# Sub-Nanosecond Electrical Pulse Switching of an Easy Plane Antiferromagnetic Insulator

Justin J. Michel, Jose Flores, and Fengyuan Yang

Department of Physics, The Ohio State University, Columbus, OH 43210, USA

Abstract

Electrical switching of antiferromagnets (AFM) is critical for AFM spintronics. However, electrical pulse-induced Neel vector reorientation in AFM insulators, while predicted to occur at much faster timescales than ferromagnetic switching, has only been demonstrated in the quasi-DC regime. Here we report reliable current-induced AFM switching in Pt/$\alpha$-$Fe_2O_3$ bilayers using electrical pulses with various durations spanning three orders of magnitude down to 0.3 ns. Together with COMSOL simulations of temperature distributions in our samples for various pulse widths, our results suggest that thermally-assisted spin-orbit torque likely play an important role for sub-ns pulses. This work demonstrates the viability of electrical switching of AFM spins using sub-ns pulses.

In recent years, research in spintronics has focused increasingly on AFM insulators due to their potential advantages, such as lack of stray magnetic fields, low damping, ultrafast dynamics, and electrical switching driven by spin-orbit torques (SOT).[1-11] Despite experimental reports of electrical switching of the Néel vector in heavy-metal (HM)/AFM-insulator bilayers, essentially all of the prior works are at DC to μs timescales, which points to the importance of investigating electrical switching of AFMs using ns or shorter pulses, as shown for their ferromagnetic counterparts.[12-17] Previously, AFM switching, magnon generation, and spin pumping have been demonstrated using ultrafast laser pulses.[17-23] However, the mechanisms of AFM switching remain elusive.[24-30] There is an ongoing debate on whether the observed electrical switching of AFM insulators is primarily driven by the thermally-induced magnetoelastic effect (ME), which requires Joule heating, or by SOT, which is expected to occur at picosecond (ps) to sub-nanosecond (ns) timescales.[26,27,31-33] Meanwhile, thermally-assisted SOT switching may also play a role, similar to ferromagnetic systems.[34-37] This is further complicated by the presence of nonmagnetic switching-like signals due to resistance change of the HM layer.[38-40] In this sense, AFM spintronics is largely dependent on demonstration of Neel vector manipulation on short timescales.

AFM insulator $\alpha$-$Fe_2O_3$ has generated much interest in this field due to its low anisotropy, high Néel temperature, and a low spin-flop transition field ($\mu_0 H < 1$ T) which allows control of its AFM spins by a magnetic field, in addition to electrical pulses. This ability is essential for our sub-ns electrical switching due to the required impedance matching (50 Ω) for delivery of such ultrashort pulses. In this letter, we report reliable electrical switching of Néel order in Pt/$\alpha$-$Fe_2O_3$ bilayers using electrical pulses with various durations down to 0.3 ns, which is our instrument limit. To our knowledge, this work presents the first experimental demonstration of electrical switching in an AFM insulator using sub-ns pulses.

We grow $\alpha$-$Fe_2O_3$(15 nm) epitaxial films on $Al_2O_3$ (0001) substrates using off-axis sputtering.[28,41] Then, a 4-nm Pt layer is deposited on $\alpha$-$Fe_2O_3$ at room temperature using off-axis sputtering. For ultrashort pulse measurements, we fabricate a Hall cross by patterning the Pt layer into an electrical switching channel (1.5 μm wide) along $Al_2O_3[1\bar{1}00]$ with two 0.5 μm wide Hall electrodes using electron-beam lithography and reactive-ion etching. Next, we pattern and deposit two 100-nm thick Ag contacts onto the two ends of the current channel with a 1.5 μm gap, resulting in a 1.5 × 1.5 $\mu m^2$ square active region, which reduces the channel resistance to 56 Ω. A scanning electron microscopy image and a schematic of the Hall cross are shown in Fig. 1a.

To deliver sub-ns electrical pulses to our device, we use the circuit shown in Fig. 1a. A fast rise-time (70 ps) voltage pulse generator with a minimum pulse width ($\Delta t$) of 0.3 ns sends an ultrashort pulse current ($\boldsymbol{I}_P$) for switching. A DC current source sends a sensing current ($\boldsymbol{I}_s$) for Hall measurement detected by a nanovoltmeter. Both the pulse generator and DC current sourcemeter are connected to an RF switch box, which can selectively connect either one to a coplanar waveguide (CPW) with a 50 Ω impedance wire-bonded to our device, such that the device forms a good termination for the transmission line (see Supplementary Information for details). All measurements in this work are performed at room temperature.

We first check the circuit's pulse delivery using an output voltage of 1 V with varying widths from a pulse generator, which is delivered through the CPW to the device. The pulses at the device are measured by an oscilloscope as shown in Fig. 1b, which exhibits a modest attenuation with a measured voltage of 0.86 V for $\Delta t \geq 20$ ns and down to 0.72 V for $\Delta t = 0.3$ ns due to imperfect impedance matching. These attenuation factors will be used for the calculation of current densities below. Since the Hall leads have significantly higher resistance (~2 kΩ), the pulse current spreading into these leads is negligible. The pulse current is calculated from $V/R$, where $V$ is the product of the output pulse voltage from the pulse generator and the attenuation factor measured in Fig. 1b, and $R = 56\ \Omega$.

To demonstrate magnetic field control of the AFM states in our device, we measure the in-plane angular dependence ($\alpha$) of transverse spin Hall magnetoresistance[42,43] (TSMR), $R_{xy}$. This effect arises from spin Hall effect-induced spin current absorption by the AFM as a function of the Néel vector orientation, manifesting as a transverse resistance modulation in the Pt layer. Figure 1c shows the $\sin 2\alpha$ angular dependence of $R_{xy}$ at $\mu_0 H \geq 0.3$ T, indicating that at a moderate magnetic field, the AFM spins in $\alpha$-$Fe_2O_3$ are almost fully aligned perpendicularly to the magnetic field in the (0001) plane via the spin-flop transition. Thus, our pulse measurements described below utilize a 0.5 T magnetic field to initialize the alignment of AFM spins.

Typical AFM switching experiments rely on multiple current channels for electrical alignment of the Néel vector along different directions.[24,27,28,44] However, such multi-channel structures with a thin Pt layer exhibit resistances much larger than 50 Ω as required for delivery of ns pulses. Here, by taking advantage of field control of AFM spins in $\alpha$-$Fe_2O_3$ films, we fabricate a Pt/$\alpha$-$Fe_2O_3$ Hall cross with ~50 Ω impedance, where the Néel vector can be aligned by a 0.5 T field to a desired orientation before electrical switching with ns pulses. In addition, the current density in our channel is more uniform than those with multiple current channels.[24,26-28]

Electrical pulse measurements are performed on our Pt/$\alpha$-$Fe_2O_3$ Hall cross with the following steps. We first apply an in-plane magnetic field of 0.5 T at $\alpha = 30°$ to align the AFM spins in the perpendicular direction as shown in Fig. 2a, and then ramp the field down to zero. As described in our previous work,[28] this leaves the $\alpha$-$Fe_2O_3$ film in a multidomain remnant state with a preferred orientation of Néel vector along $\alpha = 120°$. Subsequently, we measure TSMR ($R_{xy}$) 5 times using a sensing current $I_s = 100$ μA along the $x$-axis with a 3 s wait in between, which gives a stable $R_{xy} \approx 4$ mΩ as shown in Fig. 2b (blue points). Next, we send an electrical pulse through the current channel, wait 10 s, and measure $R_{xy}$ again (red points). Such pulse-measure sequence is performed 5 times with a 3 s wait between sequencies. Finally, a 0.5 T in-plane field is applied again at $\alpha = 30°$ to realign the AFM spins along $\alpha = 120°$ and then ramped back down to zero before the next set of measurements (blue points). This procedure is repeated multiple times for each pulse width, resulting in a clear step-like switching signal over a wide range of pulse widths of $\Delta t = 0.3$, 1, 5, and 50 ns with current densities ($J_p$) of $7.9 \times 10^{12}$, $6.6 \times 10^{12}$, $4.8 \times 10^{12}$, and $3.0 \times 10^{12}$ A/m$^2$, respectively, as shown in Figs. 2b-2e.

As observed in previous reports, the pulse application tends to rotate the Néel vector towards the $x$-axis (parallel to the pulse current), which is expected to result in a negative $R_{xy}$ change from the field-aligned state (blue points) to the pulse-switched state (red points). This pulse switching corresponds to the field-controlled AFM spin rotation in Fig. 1c from $\alpha = 30°$ to $90°$ ($\alpha = 90°$ results in AFM spins aligned along $x$-axis), which exhibits a saturated $\Delta R_{xy} = -16$ mΩ. As a comparison, the measured remnant-state $\Delta R_{xy} = -4$ mΩ is a fraction of the saturated value of -16 mΩ. This suggests that the remnant AFM state is in multiple domains with a preferred spin orientation after the application of either a magnetic field or an electrical pulse. A clear understanding of the domain configurations before and after the electrical pulses requires AFM imaging such as the X-ray magnetic linear dichroism (XMLD) photoemission electron microscopy (PEEM). Previously, XMLD-PEEM imaging of AFM domains in Pt/$\alpha$-$Fe_2O_3$ bilayers has been reported using much longer pulses.[44] For future XMLD-PEEM studies of ultrafast electrical switching of AFM films, instrumentation with impedance matching at synchrotron beamlines is needed to allow delivery of ns to sub-ns electrical pulses to samples.

The AFM switching results in step-like signals in the measured $R_{xy}$ as can be seen in Figs.

2b-2e for $\Delta t$ = 0.3 to 50 ns. In addition, nonmagnetic effects have been shown to appear in the switching measurements by multiple investigations.[39,40] This typically exhibits sawtooth-like switching signals, likely due to electromigration of grain boundaries in the thin Pt layer.

To further confirm the magnetic origin of the AFM switching driven by ultrashort electrical pulses, we reduce the applied magnetic field from 0.5 to 0.1 T and perform the pulse switching measurement with $\Delta t$ = 0.3 ns, as shown in Fig. 2f. The small field of 0.1 T is well below the spin-flop transition field of our $\alpha$-$Fe_2O_3$ films, which is unable to align the AFM spins. As a result, we observe essentially no change in the measured TSMR, indicating no switching of the Néel vector, which confirms the magnetic origin of the electrical switching.

We conduct a systematic investigation of the dependence of switching on pulse width and amplitude. For each measurement sequence with a certain pulse width, we initialize the alignment of AFM spins by an in-plane field of 0.5 T. After the field is ramped back to zero, we send pulses of the same $\Delta t$ with increasing current density and measure $R_{xy}$ after each pulse. Figure 3a shows four such sequences with $\Delta t$ = 0.3, 1, 5, and 50 ns. For each $\Delta t$ sequence, there is a threshold pulse current density ($J_{\text{th}}$), below which $R_{xy}$ remains at a high state of 4 mΩ, indicating no AFM switching. Above the $J_{\text{th}}$, $R_{xy}$ abruptly decreases to ~0 mΩ, indicating AFM switching. This demonstrates the existence of a specific threshold current density for each pulse width, and a larger $J_{\text{th}}$ is needed to switch the AFM for shorter $\Delta t$.

Figure 3b shows the obtained $J_{\text{th}}$ as a function of $\Delta t$ from 0.3 to 200 ns, where each data point is the average of 5 measurements. As $\Delta t$ increases by almost three orders of magnitude, $J_{\text{th}}$ only increases by 2.7× (from $2.6 \times 10^{12}$ A/m$^2$ at 200 ns to $7.1 \times 10^{12}$ A/m$^2$ at 0.3 ns). Figure 3c shows the $J_{\text{th}}$ vs. $1/\Delta t$ plot, which exhibits a linear dependence for $\Delta t \leq 1$ ns. Such linear dependence has been reported in SOT switching of ferromagnets from the conservation of angular momentum when thermal assistance is negligible.[12-15] We fit Fig. 3c by $J_{th} = J_{th0} + Q/\Delta t$, from which we extract $J_{th0} = 5.5 \times 10^{12}$ A/m$^2$ and Q = $4.9 \times 10^2$ C/m$^2$, where $J_{th0}$ is the minimum pulse current density needed for switching and Q is the effective charge parameter that describes the efficiency of charge to angular momentum transfer. At $\Delta t \geq 2$ ns, $J_{th}$ falls below the $1/\Delta t$ behavior, which has been attributed to thermally-assisted switching in ferromagnetic SOT switching.[12-15,34-37] We caution that there is still a notable temperature rise for sub-ns pulses, although at a much smaller magnitude than longer pulses (see COMSOL simulations below). The linear dependence in Fig. 3c should be viewed as potentially consistent with SOT switching via domain wall motion, in which thermal assistance still exists for sub-ns pulses, but plays a decreasing role for shorter pulses.

As discussed in the beginning, there are two possible mechanisms responsible for the electrical switching of AFM insulators, the thermally-induced magnetoelastic effect and SOTs.[24-32] However, distinguishing these two mechanisms has been challenging. To evaluate the thermal effects on the electrical switching of our $\alpha$-$Fe_2O_3$ films, we simulate the current density distribution as well as the temporal profile and spatial distribution of temperature ($T$) in our Hall cross structure using COMSOL Multiphysics finite element simulation software. We use the pulse profiles measured by an oscilloscope (see Fig. 1b) with pulse widths from $\Delta t$ = 0.3 to 200 ns for the simulations at threshold pulse current densities using our sample and materials parameters (see Supplementary Information for details).

Figures 4a-4f show the simulated spatial distributions of maximum temperature at $\Delta t$ = 0.3, 0.5, 1, 5, 50, and 200 ns near the end of each pulse, with an ambient temperature of 300 K, where the temperature maps exhibit some variations within the square active region of the Hall cross. The simulated maximum temperature rises ($\Delta T = T - 300$ K) at the center of the square and near a corner of a Hall lead are shown in Fig. 4g as a function of $\Delta t$ from 0.3 to 200 ns. The center of the Hall cross exhibits a relatively consistent $\Delta T$ of 100-113 K for $\Delta t > 10$ ns, and accelerated decrease of $\Delta T$ below 10 ns, which reaches 48 K at $\Delta t = 0.3$ ns. For the corner, $\Delta T$ is essentially constant within 82-88 K at $\Delta t > 1$ ns, below which it decreases to 68 K at $\Delta t = 0.3$ ns.

Lastly, we discuss the implications of the results shown in Figs. 3 and 4 regarding the potential mechanisms for the observed electrical switching of $\alpha$-$Fe_2O_3$ films. In short, it is likely that both thermal effect and SOT play a role in the electrical switching. The simulated temperature rise between 113 K for $\Delta t = 200$ ns and 48 K for $\Delta t = 0.3$ ns suggests that electrical switching of AFM spins is achieved at such elevated temperatures, for which there are two possible thermal effects. One is due to anisotropic heating in the "hot area" of the Pt layer from Joule heating, which may occur when the length/width aspect ratio of the "hot area" deviates from 1:1. The anisotropic heating induces anisotropic strain in the AFM layer, causing Néel vector switching via the thermoelastic effect. Because the "hot area" of our Hall cross is a $1.5 \times 1.5$ $\mu m^2$ square (see Fig. 1a), the temperature anisotropy between the $x$- and $y$-axis is negligible to the first order, as compared to the long channels typically used for electrical switching of AFMs in literature.[27] We note that there are subtle uneven distribution of temperature in the square "hot area" in Figs. 4a-4f due to the perturbation from the Hall leads; however, the thermoelastic effect due to these subtle variations should be much less impactful than those with "hot areas" of larger aspect ratios.

The second thermal effect is the thermally-assisted switching, where the elevated temperature lowers the barriers for AFM switching, which requires another mechanism for AFM switching such as SOTs. Figure 4g shows that the temperature rise of the Hall center decreases from 113 K at $\Delta t = 200$ ns to 48 K (58% drop) at $\Delta t = 0.3$ ns, suggesting that shorter pulses require less thermal assistance while SOTs play a more important role in AFM switching. Although the COMSOL simulations in Fig. 4 cannot conclusively confirm the roles of thermal effects, our experimental threshold current dependence on pulse width together with COMSOL simulations indicate that thermally-assisted SOT switching is the most likely mechanism for our $\alpha$-$Fe_2O_3$ films, while for sub-ns pulses, the SOTs may play a more significant role.

In summary, we achieve sub-ns electrical pulse switching of AFM spins in Pt/$\alpha$-$Fe_2O_3$ bilayers. The threshold current density varies from $2.6 \times 10^{12}$ A/m$^2$ for 200 ns pulses to $7.1 \times 10^{12}$ A/m$^2$ for 0.3 ns pulses. Our COMSOL simulations reveal temperature distributions in our Hall cross with a square "hot area". These results suggest that thermally-assisted SOTs likely play an important role for AFM switching using ultrafast electrical pulses. This work demonstrates the viability of electrical switching of AFM spins using sub-ns pulses, which is critical for AFM spintronics. Further understanding of AFM dynamics during ultrafast electrical switching process will likely require time-resolved probing techniques, which is beyond our current capabilities.

This work was supported by the Department of Energy (DOE), Office of Science, Basic Energy Sciences, under Grant No. DE-SC0001304.

**Figure Captions:**

**Figure 1.** (a) Schematic of the electrical switching circuit and device of a Pt(4 nm)/$Fe_2O_3$(15 nm) bilayer with a 1.5 × 1.5 μm$^2$ cross area and 0.5 μm wide Hall leads (inset: SEM image). The Pt layer is etched into a Hall cross and 100-nm thick silver is deposited on the pad area. (b) Time profile of pulse voltages of various widths sent from a pulse generator with a 1 V output voltage and delivered to the device circuit as measured by an oscilloscope. (c) Angular-dependent TSMR measured for a Pt(4 nm)/$Fe_2O_3$(15 nm) structure at various in-plane magnetic fields.

**Figure 2**. (a) Schematics of (left) field alignment ($\mu_0 H$ = 0.5 T, $\alpha$ = 30°) of AFM sublattice moments via spin-flop transition, and (right) electrical pulse switching of AFM moments to align along *x*-axis. (b)-(e) Electrical switching of $Fe_2O_3$ spins for various pulse widths at a pulse current density ($J_P$) above the threshold. The measurement series is as follows: First, an in-plane magnetic field is ramped up to 0.5 T and back to zero. Then, TSMR is measured repeatedly for five times (blue points). Next, before each of the five red points, a current pulse is sent through the sample, followed by a TSMR measurement. Subsequently, the magnetic field is again applied before the next set of blue points to realign the Néel vector. Any linear resistance drift is subtracted. (f) Electrical switching measurements with the same pulse parameters as in (b) for $\Delta t$ = 0.3 ns but with a lower alignment field of 0.1 T, which shows no detectable TSMR change.

**Figure 3**. (a) Dependence of TSMR on pulse current density for various pulse widths. Before each trace the AFM spins are aligned with a 0.5 T magnetic field, after which pulses of increasing current density are sent through the sample and TSMR is measured after each pulse. (b) Dependence of threshold switching current density on pulse width, where the threshold current density is defined as the pulse current density that induces $|\Delta R_{xy}|$ is > 3 mΩ in (a). Error bars are calculated from the standard deviation of 5 measurements per point. (c) Dependence of threshold

current density on $1/\Delta t$. The region of $1/\Delta t \geq 1$ GHz ($\Delta t \leq 1$ ns) is fitted with $J_{th} = J_{th0} + Q/\Delta t$ to show the linear dependence in this regime, where $J_{th0} = 5.5 \times 10^{12}$ A/m$^2$ and $Q = 4.9 \times 10^2$ C/m$^2$.

**Figure 4**. COMSOL simulation results of spatial distribution of maximum temperature near the end of each electrical pulse in a Pt/$Fe_2O_3$ Hall cross structure for various pulse widths of (a) $\Delta t$ = 0.3 ns, (b) $\Delta t$ = 0.5 ns, (c) $\Delta t$ = 1 ns, (d) $\Delta t$ = 5 ns, (e) $\Delta t$ = 20 ns, and (f) $\Delta t$ = 200 ns at the threshold pulse current densities (ambient temperature: 300 K). (g) Simulated maximum temperature rise as a function of the pulse width from 0.3 to 200 ns at the threshold current densities for both the center of the Hall cross and the corner near a Hall lead, which are marked by the solid circle and up triangle, respectively, in (a).

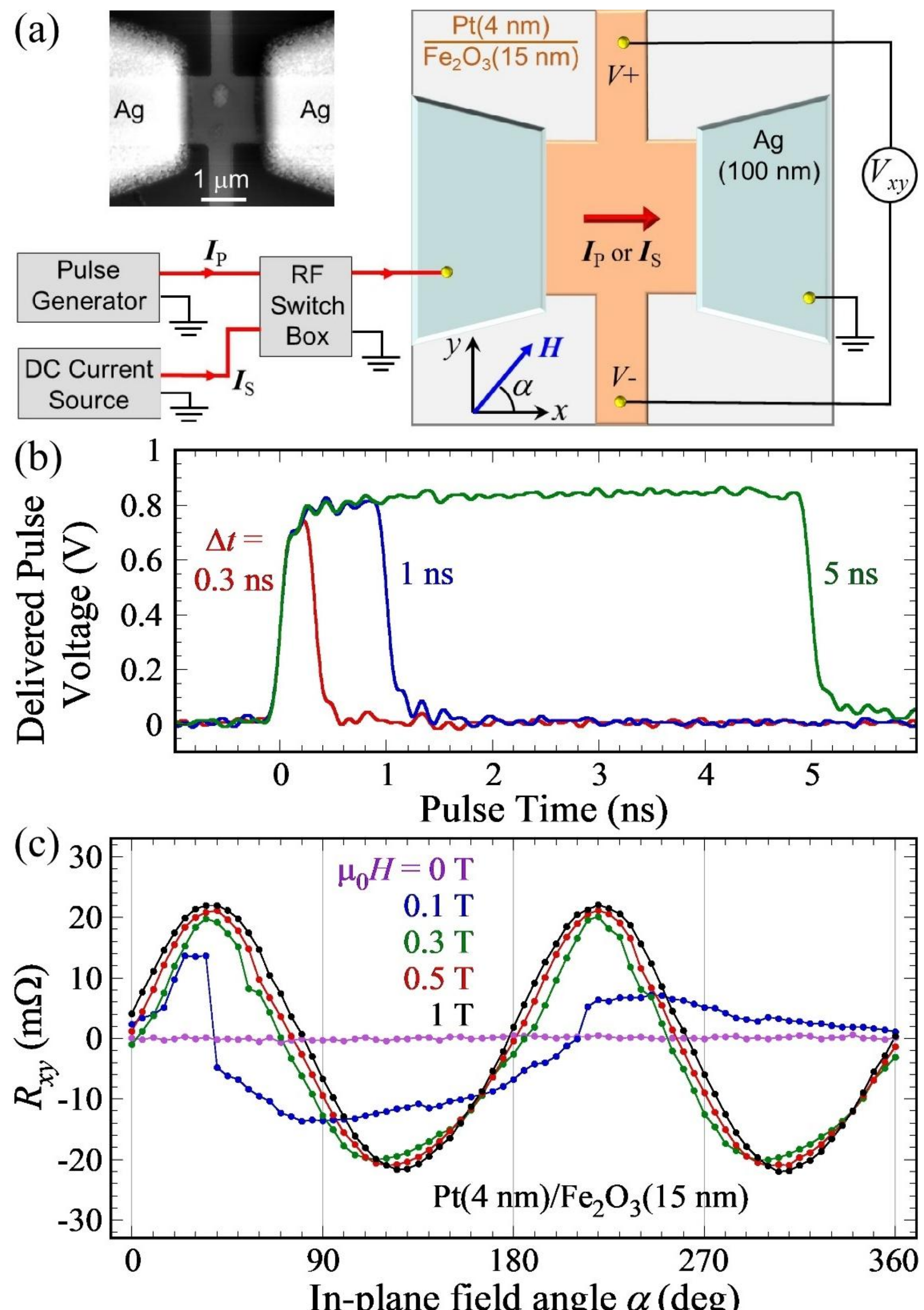

**Figure 1.**

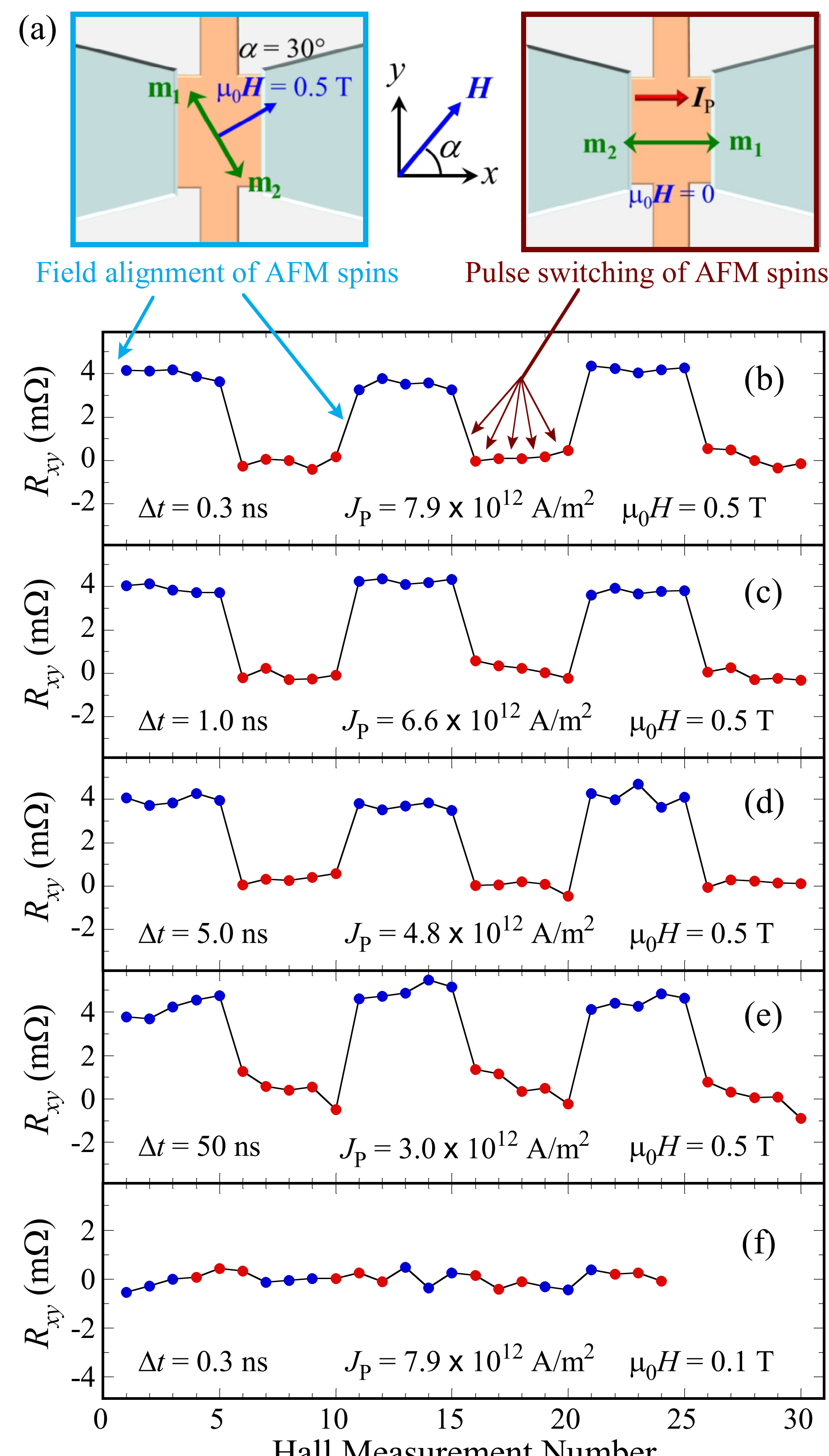

**Figure 2**.

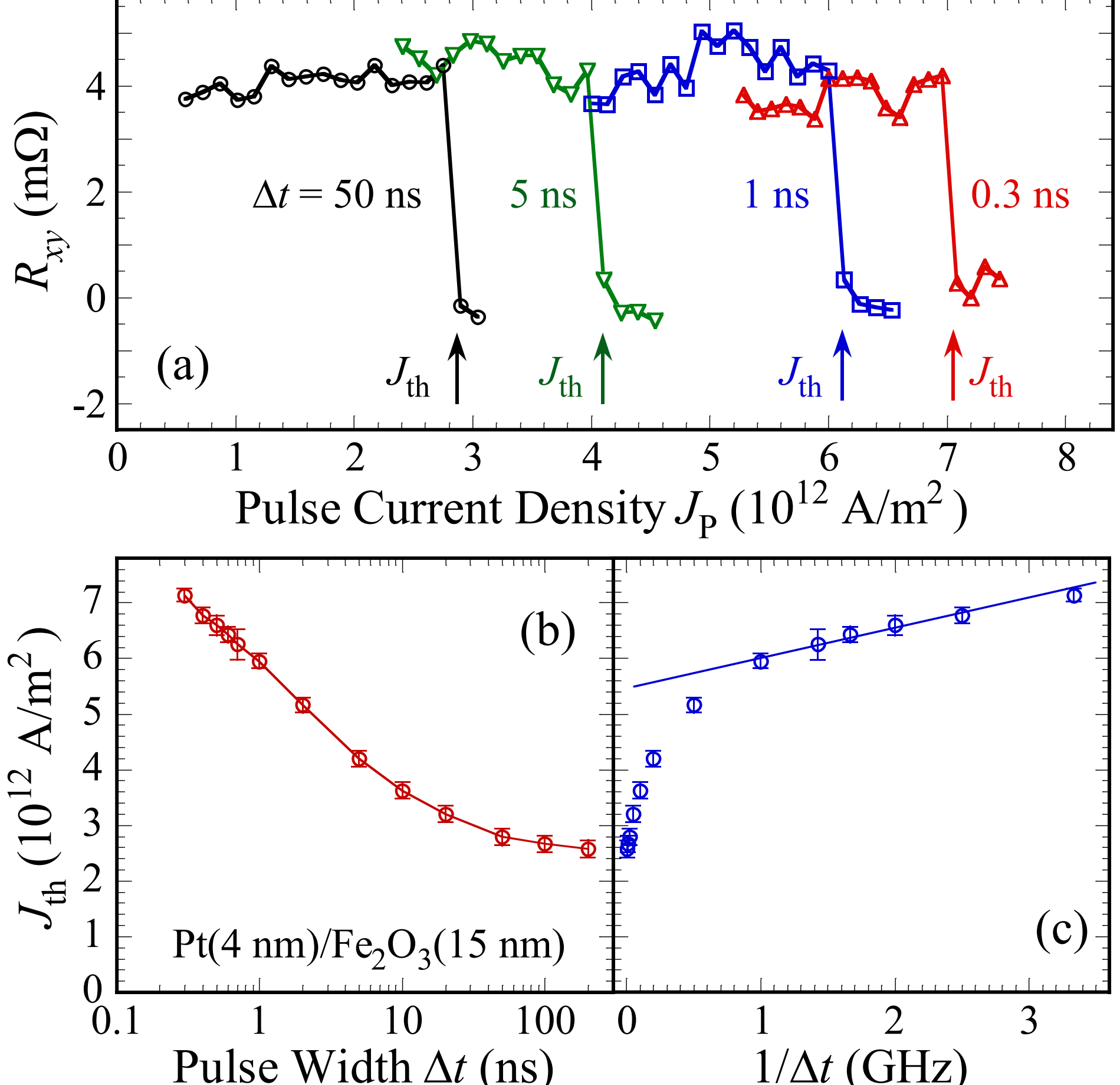

**Figure 3**.

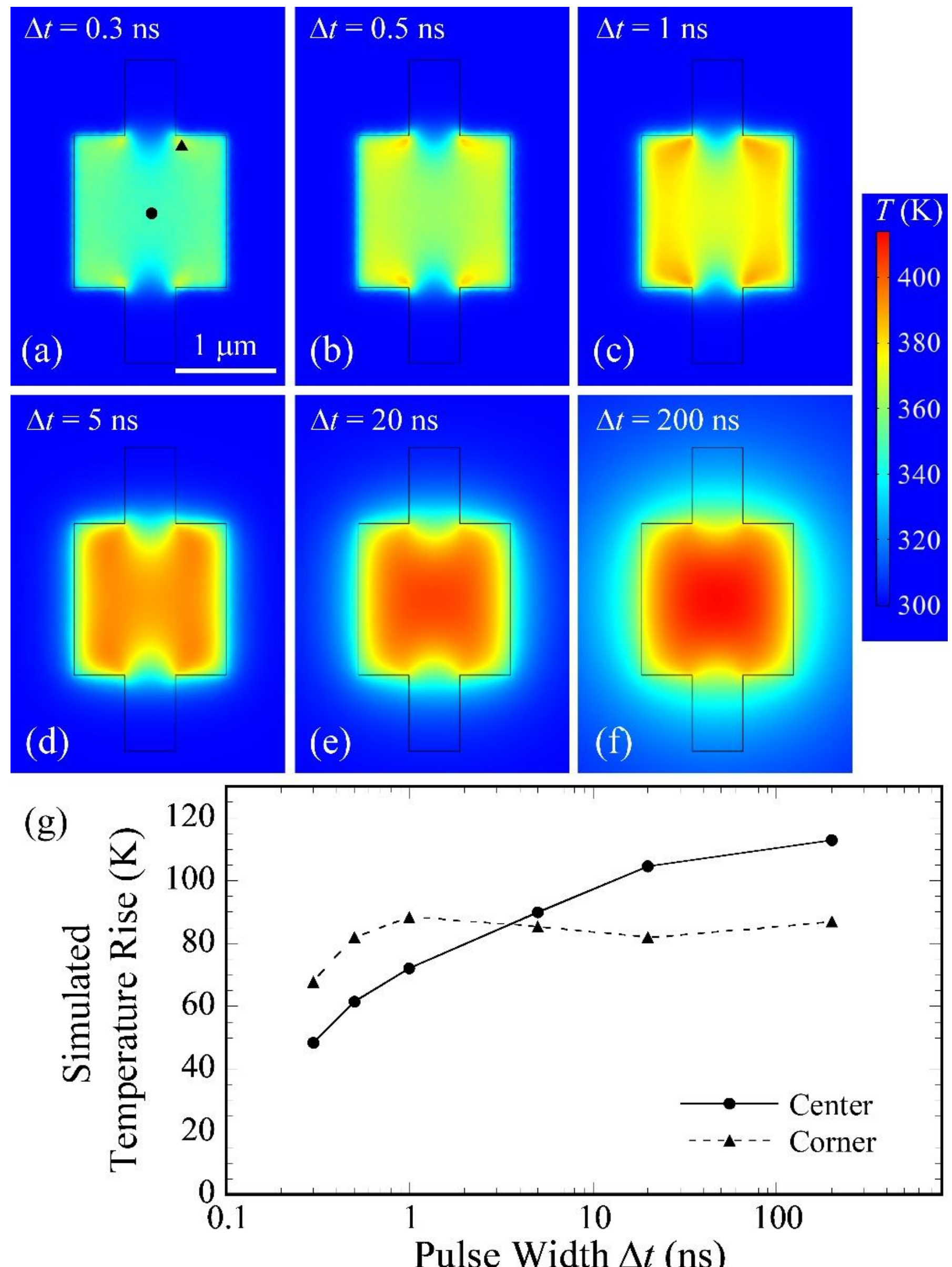

**Figure 4**.